\documentclass[aps,prc,twocolumn, superscriptaddress,showkeys,nofootinbib]{revtex4-1}  

\usepackage[utf8]{inputenc}

\usepackage{graphicx,color}
\usepackage{amsmath,amssymb,amsfonts}
\usepackage{hyperref}

\usepackage{xspace}	

\newcommand{\pcc}{\mbox{$\rho(v^{2}_{n},[p_{T}])$}\xspace}
\newcommand{\pcct}{\mbox{$\rho(v^{2}_{2},[p_{T}])$}\xspace}

\newcommand{\roots} {\mbox{$\sqrt{\textit{s}_{NN}}$}\xspace}

\def  \etas      {\mbox{$\eta / \textit{s}$ }\xspace}

\def \sc23  {\mbox{$\mathrm{SC}(2,3)$   }\xspace}
\def \sc24  {\mbox{$\mathrm{SC}(2,4)$   }\xspace}

\def \nsc23 {\mbox{$\mathrm{NSC}(2,3)$}\xspace}
\def \nsc24 {\mbox{$\mathrm{NSC}(2,4)$}\xspace}

\begin{document}
\title{Model study of the energy dependence of the correlation between anisotropic flow and the mean 
transverse momentum in Au+Au collisions
}
\medskip
\author{Niseem~Magdy} 
\email{niseemm@gmail.com}
\affiliation{Department of Physics, University of Illinois at Chicago, Chicago, Illinois 60607, USA}
\author{Petr~Parfenov}
\affiliation{National Research Nuclear University MEPhI, Moscow 115409, Russia}
\author{Arkadiy.~Taranenko}
\email{arkadij@rcf.rhic.bnl.gov}
\affiliation{National Research Nuclear University MEPhI, Moscow 115409, Russia}
\author{Iurii~Karpenko}
\affiliation{Faculty of Nuclear Sciences and Physical Engineering, Czech Technical University in Prague,  B\v rehov\'a 7, 11519 Prague 1, Czech Republic}
\author{Roy~A.~Lacey}
\email{roy.lacey@stonybrook.edu}
\affiliation{Depts. of Chemistry \& Physics, State University of New York, Stony Brook, New York 11794, USA}

\begin{abstract}
A hybrid model that employs the hadron-string transport model UrQMD and the (3+1)D relativistic viscous hydrodynamic code vHLLE, is used to investigate the beam energy dependence of the correlation coefficient $\rho(v^{2}_{2},[p_{T}])$ between the average transverse momentum $[p_{T}]$ of hadrons emitted in an event and the square of the anisotropic flow coefficient $v_2^2$. For Au+Au collisions, the model predicts characteristic patterns for the energy and event-shape dependence of the variances for $[p_{T}]$ and $v_n^2$ (${\rm Var}([p_T])$ and ${\rm Var}(v_2^2)$), and the covariance of $v_n^2$ and $[p_{T}]$ (${\rm cov}(v_{2}^{2},[p_T])$), consistent with the attenuation effects of the specific shear viscosity $\eta/s$. In contrast, $\rho(v^{2}_{2},[p_{T}])$ is predicted to be insensitive to the beam energy but sensitive to the initial-state geometry of the collisions. These observations suggest that a precise set of measurements for ${\rm Var}([p_T])$,  ${\rm Var}(v_2^2)$,  ${\rm cov}(v_{2}^{2},[p_T])$ and  $\rho(v^{2}_{2},[p_{T}])$ as a function of beam-energy and event-shape, could aid precision extraction of the temperature and baryon chemical-potential dependence of  $\eta/s$ from the wealth of Au+Au data obtained in the RHIC beam energy scan.
\end{abstract}
\maketitle

\section{Introduction}
The collision of heavy nuclei at ultra-relativistic energies results in the creation of  matter known as 
the quark-gluon plasma (QGP) \cite{Shuryak:1978ij,Shuryak:1980tp,Muller:2012zq}.
A key aim of the heavy-ion programs at the Large Hadron Collider (LHC) and the Relativistic Heavy-Ion Collider (RHIC) is to chart the transport properties of the QGP as a function of temperature ($T$) and baryon chemical potential ($\mu_{B}$). Over the last few years, considerable theoretical and experimental attention have been given to the extraction of the specific shear viscosity -- the ratio of shear viscosity to entropy density $\etas(T, \mu_{B})$. This ratio characterizes the QGP's ability to transport momentum and is commonly termed a final-state effect.
Anisotropic flow ($v_n$) measurements, made as a function of collision centrality and transverse momentum ($p_{T}$), have played a central role in such extractions. This is because they constrain the viscous hydrodynamic response to the spatially-anisotropic energy-density distribution produced in the initial stages of the collision~\cite{Danielewicz:1998vz,Ackermann:2000tr,Adcox:2002ms,Heinz:2001xi,Huovinen:2001cy,Hirano:2002ds,Shuryak:2003xe,Hirano:2005xf,Romatschke:2007mq,Luzum:2008cw,Bozek:2009dw,Song:2010mg,Qian:2016fpi,Schenke:2011tv,Teaney:2012ke,Gardim:2012yp,Lacey:2013eia,Magdy:2020gxf}.

Precision extraction of  $\etas(T, \mu_{B})$ via model comparisons to flow measurements require robust constraints for the requisite estimate of the initial-state employed in the model calculations \cite{Bernhard:2016tnd}. Such estimates have not been fully charted quantitatively. Recently, several works have sought to study the correlation coefficient \pcc because it indicates different sensitivities to the initial- and final-states of the collisions~\cite{Bozek:2016yoj,Giacalone:2017uqx,Giacalone:2020byk,Bozek:2016yoj,Bozek:2020drh,Schenke:2020uqq,Giacalone:2020dln,Lim:2021auv,ATLAS:2021kty,Magdy:2021ocp} that could provide more stringent constraints for $\etas(T, \mu_{B})$. The correlation coefficient 
\begin{eqnarray}
    \rho(v^{2}_{n},[p_{T}]) = \frac{{\rm cov}(v_{n}^{2},[p_T])}{\sqrt{{\rm Var}(v_{n}^{2})} \sqrt{{\rm Var}([p_T])}}\label{eq:1},
\end{eqnarray}
reflects the correlation between the n$^{\rm th}$-order flow harmonics $v_{n}$ and the average transverse momentum of the particles in an event $[p_{T}]$. The $p_T$-fluctuations result from fluctuations in the initial size of the fireball \cite{Broniowski:2009fm,Bozek:2012fw}.
Thus, correlations between the average transverse radial flow and the $v_n$ coefficients could encode 
crucial information on (i) the correlation between the size and the eccentricities in the initial state, and (ii) on the correlations of the  strength of the hydrodynamic response with the flow coefficients.
Note that $v_{n}$ is eccentricity-driven while the $[p_{T}]$ is transverse-size-driven; that is, events with similar energy-density but smaller transverse-size [of the collision zone], generate more radial expansion and consequently, larger $[p_{T}]$~\cite{Bozek:2012fw}. The correlation coefficient has also been shown to be sensitive to the correlations between the initial-size and the initial-state deformation of the colliding nuclei~\cite{Giacalone:2019pca,Giacalone:2020awm,ATLAS:2021kty}.

Measurements of the  \pcct correlation coefficient have been reported for p+Pb and Pb+Pb collisions at  \roots = 5.02~TeV~\cite{Aad:2019fgl}. They indicate characteristic patterns which could provide important model constraints for the initial-state in these collisions. Similar constraints are anticipated from the wealth of measurements for different systems and beam energies underway at RHIC.

The correlation coefficient has also been investigated in hydrodynamic and transport models~\cite{Bozek:2021zim,Bozek:2020drh,Giacalone:2020dln,Lim:2021auv,Magdy:2021ocp,Giacalone:2021clp}. However, a detailed mapping of its beam energy dependence is still lacking. Such a mapping could be invaluable for model comparisons to the wealth of measurements from the RHIC beam energy scan, resulting in improved constraints for the extraction of $\etas(T, \mu_{B})$.
\begin{table*}[t]
\begin{center}
 \begin{tabular}{|c|c|c|c|c|}
 \hline 
  Model          &~\etas~&~Initial state                 &          \roots           \\
  \hline
  AMPT           &~0.1~  &~HIJING model                  &          200~GeV          \\
 \hline 
  EPOS           &~0.08~ &~Flux tubes using Gribov-Regge &          200~GeV          \\
       $ $       &$ $    & multiple scattering theory    &          $ $              \\
 \hline 
  hybrid  &~0.08~ &~UrQMD model                   &          200~GeV          \\
 \hline 
  hybrid  &~0.08~ &~UrQMD model                   &          62.5~GeV         \\
 \hline 
  hybrid  &~0.12~ &~UrQMD model                   &          27.0~GeV         \\
 \hline 
  hybrid  &~0.15~ &~UrQMD model                   &          19.5~GeV         \\
 \hline 
\end{tabular}
\caption{ A summary of the models used [in this, and a prior study] and several of their associated parameters.}
\label{tab:1}
\end{center}
\end{table*}
  
 Here, we use simulated events from a hybrid model~\cite{Karpenko:2015xea} that employs the hadron-string transport model UrQMD and the (3+1)D relativistic viscous hydrodynamic code vHLLE, to chart the beam-energy and even-shape dependence of $\rho(v^{2}_{2},[p_{T}])$ and its components (${{\rm Var}(v_{n}^{2})}, {\rm Var}([p_T])$ and ${\rm cov}(v_{n}^{2},[p_T])$) in Au+Au collisions. The model studies, which incorporate the effects of both the initial- and final-state, compliments and extends our earlier study of $\rho(v^{2}_{2},[p_{T}])$ in Au+Au collisions at \roots = 200~GeV, with the AMPT and EPOS models~\cite{Magdy:2021ocp}.

The paper is organized as follows.
Section~\ref{sec:2} summarizes the theoretical model used to investigate the  beam energy dependence $\rho(v^{2}_{2},[p_{T}])$ and the details of the analysis method employed. The results from the model studies are presented in Sec.~\ref{sec:3} followed by a summary in Sec.~\ref{sec:4}.

\section{Methodology} \label{sec:2}

\subsection{Models}\label{sec:2a}

The simulated events were obtained with a hybrid model~\cite{Karpenko:2015xea} that employs the hadron-string transport model UrQMD~\cite{Bleicher:1999xi,Bass:1998ca} for the early and late non-equilibrium stages of the collision, and the (3+1)D relativistic viscous hydrodynamic code vHLLE~\cite{Karpenko:2013wva, Karpenko:2015xea} for the quark-gluon plasma phase. The latter uses an equation of state based on the Chiral model (XPT EoS), which has a crossover transition between the QGP and hadronic phases for all baryon densities. Fluid to particle transition, or particlization, is performed when the energy density $\epsilon$ in the hydro cells reaches the switching value $\epsilon_{SW}$ = 0.5~GeV/$fm^3$. The UrQMD hadronic cascade serves to generate the hadronic re-scatterings and decays. The initial state parameters, hydrodynamic starting time $\tau_{0}$  and $\eta/s$ in the fluid phase are tuned for each collision energy to reproduce the experimental bulk observables such as the (pseudo)rapidity distributions, transverse momentum spectra, and the elliptic flow coefficient for inclusive charged hadrons. Further details of the model and the model parameters are detailed in Ref.~\cite{Karpenko:2015xea}.


Events were generated for Au+Au collisions spanning a broad set of centralities for each beam energy selected in the range $\sqrt{s_{NN}}$ = 19.5 -- 200~GeV. The observables were evaluated for charged hadrons with $0.2 < p_T < 2.0$ GeV/$c$ and pseudorapidity $|\eta| < 1.0$. The latter selection mimics the acceptance of the STAR experiment at RHIC. The new results obtained with the hybrid model are also compared  to those obtained in our earlier work~\cite{Magdy:2021ocp} which used the AMPT~\cite{Lin:2004en}, and EPOS~\cite{Drescher:2000ha,Werner:2010aa,Werner:2013tya} models. Table~\ref{tab:1} gives a summary of the models used and their associated parameters.

\subsection{Analysis Method}\label{sec:2b}
The experimental  \pcct correlation coefficient (Eq.~\ref{eq:1})  involves the evaluation of variances and covariances that employ two- and multi-particle correlations. These correlations could be subject to non-flow effects resulting from resonance decays, Bose-Einstein correlations, and the fragments of individual jets~\cite{Jia:2013tja}. However, such non-flow effects are dominated by particles emitted within a localized $\mathrm{\eta}$-region and can be minimized using the subevent cumulant method~\cite{Jia:2017hbm,Huo:2017nms,Zhang:2018lls,Magdy:2020bhd}. The efficacy of the method has been demonstrated for two- and multi-particle correlations~\cite{Jia:2017hbm,Huo:2017nms,Magdy:2020bhd}.

In this study, we used the two-subevents method to evaluate the variance of  $v_{2}^{2}$.
We use two separate $\eta$ ranges defined as $-1.0< \eta_{A}<-0.35$  and $0.35<\eta_{C}<1.0$ to obtain the $v_{2}^{2}$ variance as:
\begin{eqnarray}\label{eq:2-1}
    {\rm Var}(v_{2}^{2}) &=& v_{2}\{2\}^{4} - v_{2}\{4\}^{4} \nonumber,\\
     &=& C^{2}_{2}\{2\} - C_{2}\{4\},
\end{eqnarray}
where $v_{2}\{2\}$ and $v_{2}\{4\}$ are the two- and four-particle elliptic flow correlations from the subevent method~\cite{Jia:2017hbm},
\begin{eqnarray}\label{eq:2-2}
C_{2}\{2\} &=&  \langle \langle 2\rangle\rangle|_{A,C}   =  \langle  \langle e^{\textit{i}~ 2 (\varphi^{A}_{1} -  \varphi^{C}_{2} )} \rangle \rangle,\\
v_{2}\{2\}     &=&  \sqrt{C_{n}\{2\}}
\end{eqnarray}
where $\phi_{A(C)}$ is the azimuthal angle of particles in the range $\eta_{A}$ ($\eta_{C}$).
\begin{eqnarray}\label{eq:2-3}
C_{2}\{4\}    &=&  \langle \langle 4\rangle\rangle|_{A,C} - 2 \langle \langle 2\rangle\rangle^{2}|_{A,C},\\
v_{2}\{4\}     &=& \sqrt[4]{-C_{2}\{4\} } 
\end{eqnarray}
where,
\begin{eqnarray}\label{eq:2-4}
\langle \langle 4\rangle\rangle|_{A,C} &=&  \langle  \langle e^{\textit{i}~ 2 (\varphi^{A}_{1} + \varphi^{A}_{2} -  \varphi^{C}_{3} -  \varphi^{C}_{4})} \rangle \rangle.
\end{eqnarray}

The variance of the mean $p_{T}$, $c_k \sim {\rm Var}([p_T])$~\cite{Abelev:2014ckr}, computed in the range $|\eta_{B}|<0.35$, can be given as:
\small{
\begin{eqnarray}\label{eq:2-5}
    c_k  = \left\langle  \frac{1}{N_{\rm pair}} \sum_{B}\sum_{B^{\prime}\neq B} (p_{T,B} - \langle [p_T] \rangle )  (p_{T,B^{\prime}} - \langle [p_T] \rangle) \right\rangle,  \nonumber \\
\end{eqnarray}
}
where  $\langle  \rangle$ is an average over all events. The event mean $p_T$, ($[p_T]$)  is then given as,
\begin{eqnarray}\label{eq:2-6}
     [p_T]  =  \sum^{M_{B}}_{i=1} p_{T,i}  /  M_{B},
\end{eqnarray}
where $M_{B}$ is the number of tracks in subevent $B$.

The  covariance   of $v_{2}^{2}$ and the $[p_T]$ (${\rm cov}(v_{2}^{2},[p_T])$) is obtained via the three-subevents method~\cite{Aad:2019fgl,Zhang:2021phk} as,
\small{
\begin{eqnarray}\label{eq:2-7}
{\rm cov}(v_{2}^{2},[p_T]) &=&  {\rm Re} \left( \left< \sum_{A,C} e^{i2(\phi_{A} - \phi_{C})} \left( [p_T] - \langle [p_T] \rangle \right)_{B} \right> \right). \nonumber \\
\end{eqnarray}
}
The $\rho(v^{2}_{2},[p_{T}])$ correlation coefficient~\cite{Giacalone:2020byk,Lim:2021auv,Bozek:2016yoj,Bozek:2020drh,Schenke:2020uqq,Giacalone:2020dln,ATLAS:2021kty} is then obtained via Eqs.~\ref{eq:2-1}, \ref{eq:2-5} and \ref{eq:2-7};
\begin{eqnarray}\label{eq:2-8}
    \rho(v^{2}_{2},[p_{T}]) = \frac{{\rm cov}(v_{2}^{2},[p_T])}{\sqrt{{\rm Var}(v_{2}^{2})} \sqrt{c_k}}.
\end{eqnarray}

\section{Results and discussion}\label{sec:3}

\begin{figure}[t] 
\includegraphics[width=1.0 \linewidth, angle=-0,keepaspectratio=true,clip=true]{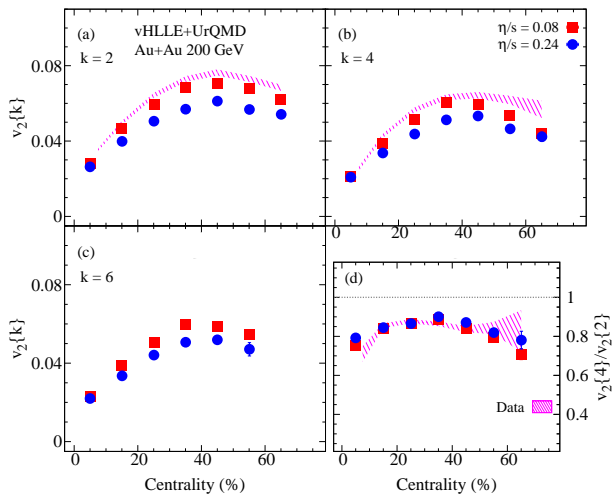}
\vskip -0.4cm
\caption{Centrality dependence of $v_{2}\lbrace 2\rbrace$ (a), $v_{2}\lbrace 4\rbrace$ (b), $v_{2}\lbrace 6\rbrace$ (c), and $v_{2}\lbrace 4\rbrace / v_{2}\lbrace 2\rbrace$  computed with the hybrid  model for Au+Au collisions at \roots = 200~GeV for \etas = 0.08 and 0.24. The hatched bands represent the experimental data reported in Ref.~\cite{Adams:2004bi}. 
  }\label{fig:1}
\vskip -0.3cm
\end{figure}

The correlation coefficient $\rho(v^{2}_{2},[p_{T}])$ is obtained from the correlations and fluctuations of $v_2$ and $p_T$ .
Consequently, it is informative to investigate the dependence of the magnitude of $v_2$ [and its fluctuations]  on (i) \etas for a selected beam energy and (ii) the beam energy for a fixed system. 
Figure~\ref{fig:1} shows a comparison of the centrality dependence of  
$v_{2}\lbrace 2\rbrace$ (a), $v_{2}\lbrace 4\rbrace$ (b) , $v_{2}\lbrace 6\rbrace$ (c) and the ratios $v_{2}\lbrace 4\rbrace/v_{2}\lbrace 2\rbrace$ (d) obtained from the  hybrid model simulations with \etas = 0.08 and 0.24.  
Panels (a), (b), and (c) show that the flow coefficients are sensitive to the \etas variations. However, panel (d) shows that the ratios $v_{2}\lbrace 4\rbrace/v_{2}\lbrace 2\rbrace$ are relatively insensitive to \etas.
The hatched bands in panels (a), (b), and (d)  which show the STAR measurements reported in Ref.~\cite{Adams:2004bi}, indicate comparable magnitudes for the experimental and simulated  data.
The ratios $v_{2}\{4\}/ v_{2}\{2\}$, give a measure of the strength of the elliptic flow fluctuations. They can be seen to vary with centrality suggesting that the flow fluctuations are eccentricity-driven and are roughly a constant fraction of $v_{2}\{2\}$.

\begin{figure}[t] 
\includegraphics[width=1.0 \linewidth, angle=-0,keepaspectratio=true,clip=true]{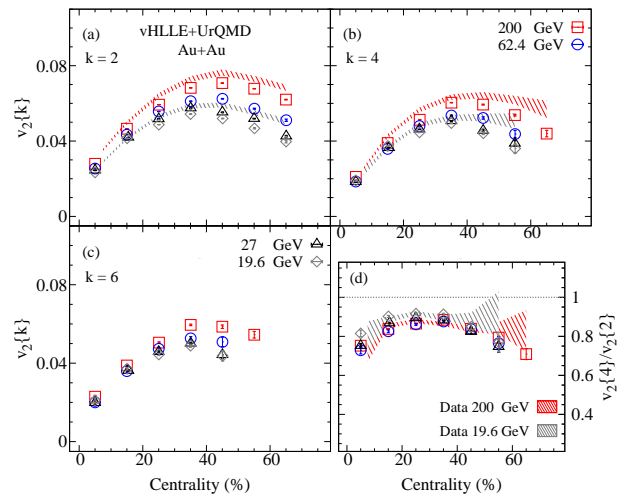}
\vskip -0.4cm
\caption{Centrality dependence of $v_{2}\lbrace 2\rbrace$ (a), $v_{2}\lbrace 4\rbrace$ (b), $v_{2}\lbrace 6\rbrace$ (c), and $v_{2}\lbrace 4\rbrace / v_{2}\lbrace 2\rbrace$  (d) computed with the hybrid model for Au+Au collisions at \roots = 200, 62.4, 27 and 19.6~GeV. The hatched bands represent the experimental data reported for \roots = 200 and 19.6~GeV in Refs.~\cite{Adams:2004bi} and \cite{STAR:2012och}.
  }\label{fig:2}
\vskip -0.3cm
\end{figure}
\begin{figure}[t]
\begin{center}
    \includegraphics[width=1.0 \linewidth, angle=-0,keepaspectratio=true,clip=true]{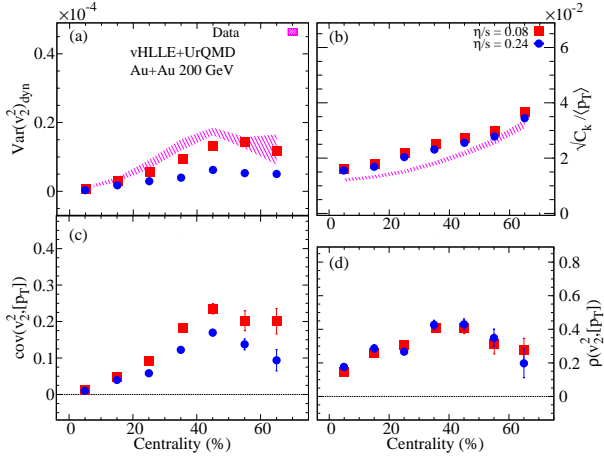}
\end{center}
      \vskip -0.8cm
    \caption{    
		Comparison of the centrality dependence of the values for ${\rm Var}(v_{n}^{2})_{dyn}$ (a), $\sqrt{c_{k}}/\langle p_{T}\rangle$ (b), ${\rm cov}(v_{n}^{2},[p_T])$ (c) and $\rho(v^{2}_{n},[p_{T}])$ (d), computed for Au+Au collisions at \roots = 200~GeV with the hybrid model.  Results are compared for \etas = 0.08 and \etas = 0.24 as indicated. The hatched bands show the values for ${\rm Var}(v_{2}^{2})$ (a) and $\sqrt{c_{k}}/\langle p_{T}\rangle$ (b) evaluated from the experimental data published in Refs.~\cite{Adams:2004bi,STAR:2019dow}. \label{fig:3}
  }
\vskip -0.3cm
\end{figure}
\begin{figure}[tb]
\begin{center}
    \includegraphics[width=1.0 \linewidth, angle=-0,keepaspectratio=true,clip=true]{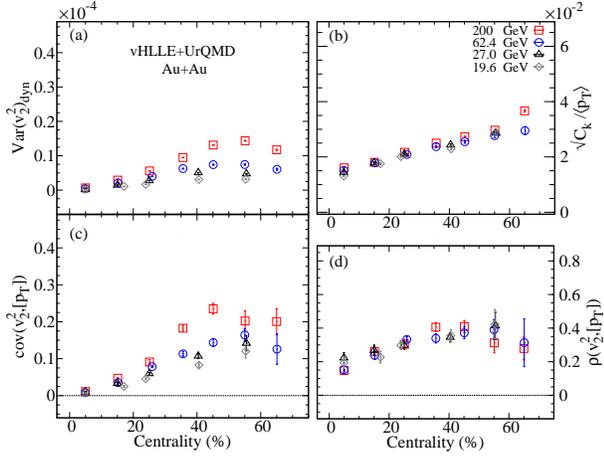}
\end{center}
      \vskip -0.8cm
    \caption{    
		Comparison of the centrality dependence of the values for ${\rm Var}(v_{n}^{2})_{dyn}$ (a), $\sqrt{c_{k}}/\langle p_{T}\rangle$ (b), ${\rm cov}(v_{n}^{2},[p_T])$ (c) and $\rho(v^{2}_{n},[p_{T}])$ (d), computed for Au+Au collisions at \roots = 200, 62.4, 27 and 19.6~GeV with the hybrid model.  \label{fig:4}
  }
\vskip -0.3cm
\end{figure}

\begin{figure}[hbt]
\begin{center}
    \includegraphics[width=1.0 \linewidth, angle=-0,keepaspectratio=true,clip=true]{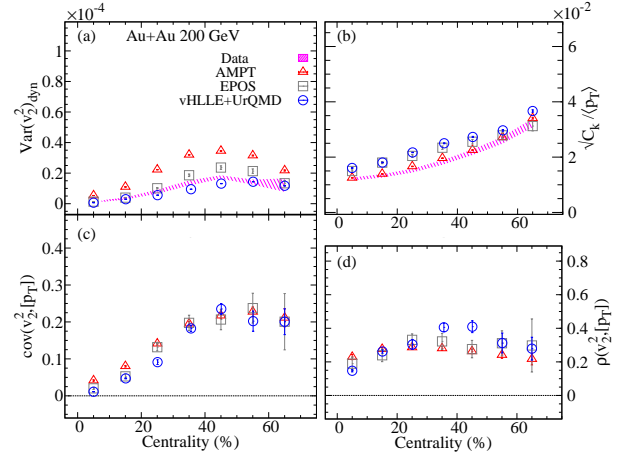}
\end{center}
      \vskip -0.8cm
    \caption{    
  Comparison of the centrality dependence of ${\rm Var}(v_{n}^{2})_{dyn}$ (a), $\sqrt{c_{k}}/\langle p_{T}\rangle$ (b), ${\rm cov}(v_{n}^{2},[p_T])$ (c) and $\rho(v^{2}_{n},[p_{T}])$ (d) obtained with the hybrid, AMPT and EPOS models for Au+Au collisions at \roots = 200~GeV. The results for the AMPT and EPOS models are taken from Ref.~\cite{Magdy:2021ocp}.  The hatched bands represent the experimental data reported  in Refs.~\cite{Adams:2004bi,STAR:2019dow}. \label{fig:5}
  }
	\vskip -0.3cm
\end{figure}
%
\begin{figure}[hbt]
\begin{center}
    \includegraphics[width=1.0 \linewidth, angle=-0,keepaspectratio=true,clip=true]{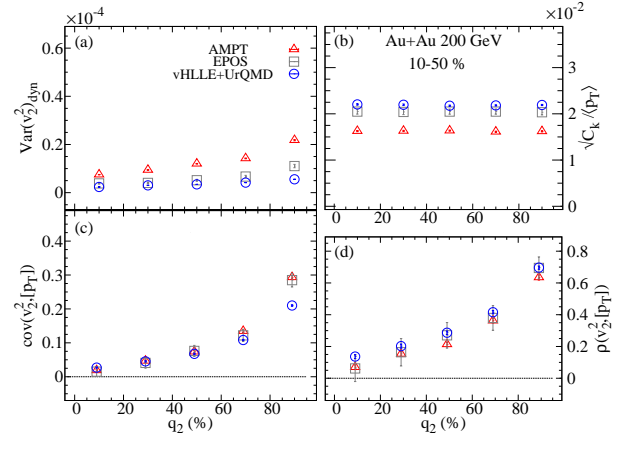}
\end{center}
      \vskip -0.8cm
    \caption{    
Comparison of the  $q_{2}$-dependent of ${\rm Var}(v_{n}^{2})_{dyn}$ (a), $\sqrt{c_{k}}/\langle p_{T}\rangle$ (b), ${\rm cov}(v_{n}^{2},[p_T])$ (c) and $\rho(v^{2}_{n},[p_{T}])$ (d) computed for  $10-50\%$ central Au+Au collisions ($\sqrt{s_{NN}}$ = 200~GeV) obtained with the hybrid, AMPT and EPOS models. The results for the AMPT and EPOS models are taken from Ref.~\cite{Magdy:2021ocp}. \label{fig:6}
  }
\vskip -0.3cm
\end{figure}

The beam-energy dependence of $v_{2}\lbrace 2\rbrace$ (a), $v_{2}\lbrace 4\rbrace$ (b) , $v_{2}\lbrace 6\rbrace$ (c) and the ratios $v_{2}\lbrace 4\rbrace/v_{2}\lbrace 2\rbrace$ (d) are shown in  Fig.~\ref{fig:2}. Panels (a), (b), and (c)  show that the experimental~\cite{Adams:2004bi,STAR:2012och} [hatched bands] and simulated flow coefficients are sensitive to the beam energy variations and follow a pattern similar to the one shown in Fig.~\ref{fig:1}. Such a pattern is to be expected if a significant change in $\etas(T, \mu_{B})$ results from a change in beam energy (cf. Tab.~\ref{tab:1}). However, it is noteworthy that the stiffness of the EoS, the duration of the pre-hydrodynamic and hydrodynamic phases, and the nuclear passage time, all change with the beam energy and could influence $v_n$. Panel (d) shows that the experimental and simulated ratios $v_{2}\{4\}/ v_{2}\{2\}$depend weakly on beam energy, suggesting that sizable variations of the beam energy do not strongly influence the flow fluctuations.

Figure~\ref{fig:3} shows the influence of a change in the magnitude of $\eta/s$ on the hybrid model results for ${\rm Var}(v_{2}^{2})$ (a), $\sqrt{c_{k}}/\langle p_{T}\rangle$ (b), ${\rm cov}(v_{2}^{2},[p_T])$ (c) and $\rho(v^{2}_{2},[p_{T}])$ (d). 
The simulated results are compared for \etas = 0.08 and 0.24 as indicated.
The hatched bands show the values for ${\rm Var}(v_{2}^{2})$ (a) and $\sqrt{c_{k}}/\langle p_{T}\rangle$ (b) evaluated from the experimental data published in Refs.~\cite{Adams:2004bi} and \cite{STAR:2019dow}.
Figs.~\ref{fig:3} (a) and (c) show that an increase in the magnitude of \etas leads to a reduction in the magnitudes of  ${\rm Var}(v_{2}^{2})$ and ${\rm cov}(v_{2}^{2},[p_T])$  over the full centrality range. By contrast,  the values for $\sqrt{c_{k}}/\langle p_{T}\rangle$ (b) and  $\rho(v^{2}_{2},[p_{T}])$ (d) show little, if any, change with \etas.  The implied insensitivity of $\rho(v^{2}_{2},[p_{T}])$ to \etas could result from a possible cancellation of the viscous attenuation effects in the \pcct correlation coefficient.
%

Figure~\ref{fig:4} shows the beam-energy dependence of ${\rm Var}(v_{2}^{2})$ (a), $\sqrt{c_{k}}/\langle p_{T}\rangle$ (b), ${\rm cov}(v_{2}^{2},[p_T])$ (c) and \pcct (d). They indicate patterns which are strikingly similar to the ones which result from a change in \etas as illustrated in Fig.~\ref{fig:3}. As discussed earlier, such a pattern would be expected if a change in the magnitude of the beam energy results in a significant change in \etas (cf. Tab.~\ref{tab:1}). These results suggest that measurements of the beam energy dependence of ${\rm Var}(v_{2}^{2})$ and  ${\rm cov}(v_{2}^{2},[p_T])$ could provide important constraints for \etas while the measurements for \pcct provide complimentary constraints for the initial-state eccentricity and its fluctations.

Figure~\ref{fig:5} compares the centrality dependence of the values for ${\rm Var}(v_{2}^{2})$ (a),  $\sqrt{c_{k}}/\langle p_{T}\rangle$ (b), ${\rm cov}(v_{2}^{2},[p_T])$ (c), and $\rho(v^{2}_{2},[p_{T}])$ (d) in Au+Au collisions ($\sqrt{s_{NN}}$ = 200~GeV) simulated with the hybrid, AMPT, and EPOS models. The comparison indicates good overall agreement between the models, suggesting comparable initial- and final-state effects in the three models for $\sqrt{s_{NN}}$ = 200~GeV. The hatched bands in panels (a) and (b) represent the values estimated from the experimental measurements reported in Refs.~\cite{Adams:2004bi} and \cite{STAR:2019dow}. They show good qualitative agreement with the corresponding simulated results as well.

The influence of the shape and size of the collision systems on $\rho(v^{2}_{n},[p_{T}])$ can be studied using the  Event Shape Engineering (ESE) technique~\cite{Adler:2002pu}. The ESE technique tests the sensitivity of the correlation coefficient to (i) the event-by-event fluctuations of the  $v_{n}$ coefficients for a fixed centrality and (ii) shape variations of the collision system~\cite{Abelev:2012di}. In the current study, the event-shape selections were conducted via fractional cuts on the reduced second-order flow vector ($q_2$) distribution~\cite{Schukraft:2012ah,Adler:2002pu};
%
\begin{eqnarray}
Q_{2, x} &=& \sum_{i} \cos(2 \varphi_{i}), \\
Q_{2, y} &=& \sum_{i} \sin(2 \varphi_{i}), \\
q_{2}    &=& \frac{\sqrt{Q_{2, x}^2 + Q_{2, y}^2}}{\sqrt{M}},
\end{eqnarray}
where $Q_{2}$ is the magnitude of the second-order flow vector determined within $\mathrm{1.5< \eta < 2.5}$, and $M$ is the multiplicity of charged hadrons in the sub-event used. Note that the correlators are constructed in subevents for $|\eta| < 1.0$ to guarantee the separation between the sub-events used to estimate  $q_{2}$ and $\rho(v^{2}_{n},[p_{T}])$. 

Figure~\ref{fig:6} compares the $q_2\%$ dependence of ${\rm Var}(v_{2}^{2})$ (a), $\sqrt{c_{k}}/\langle p_{T}\rangle$ (b), ${\rm cov}(v_{2}^{2},[p_T])$ (c) and $\rho(v^{2}_{2},[p_{T}])$ (d),  for Au+Au collisions at \roots = 200~GeV, obtained with the hybrid, AMPT and EPOS models.
The results for all three models are similar. However, they show a strikingly weak dependence for $c_k$ (panel (b))  and an essentially quadratic dependence of ${\rm cov}(v_{2}^{2},[p_T])$  and $\rho(v^{2}_{2},[p_{T}])$ on $q_2\%$. 
These dependencies suggests that ESE measurements taken in tandem with beam-energy dependent measurements, could provide new insights on initial- and final-state effects and facilitate precision extraction of $\etas (T, \mu_{B})$.


\section{Conclusion} \label{sec:4} 
In summary, we have investigated the model predictions for the beam-energy and event-shape dependence of ${\rm Var}([p_T])$, ${\rm Var}(v_2^2)$, ${\rm cov}(v_{2}^{2},[p_T])$ and $\rho(v^{2}_{2},[p_{T}])$ in Au+Au collisions, with a hybrid model that employs the hadron-string transport model UrQMD for the early and late non-equilibrium stages of the collision, and the (3+1)D relativistic viscous hydrodynamic code vHLLE for the quark-gluon plasma phase. 
The model predicts characteristic patterns for the energy and event-shape dependence of the variances for $[p_{T}]$ and $v_n^2$ (${\rm Var}([p_T])$ and ${\rm Var}(v_2^2)$), and the covariance of $v_n^2$ and $[p_{T}]$ (${\rm cov}(v_{2}^{2},[p_T])$), consistent with the attenuation effects of the specific shear viscosity $\eta/s$.
By contrast, the correlation coefficient $\rho(v^{2}_{2},[p_{T}])$ is predicted to be insensitive to $\eta/s$ but sensitive to the initial-state geometry 
of the collisions. These predictions suggest that a precise set of measurements for ${\rm Var}([p_T])$,  ${\rm Var}(v_2^2)$,  ${\rm cov}(v_{2}^{2},[p_T])$ and  $\rho(v^{2}_{2},[p_{T}])$ as a function of beam-energy and event-shape, could aid precision extraction of the temperature and baryon-chemical-potential dependence of $\eta/s$ from the Au+Au collisions obtained in the RHIC beam energy scan.

\section*{Acknowledgments}
%
%
This research is supported by the US Department of Energy, Office of Nuclear Physics (DOE NP),  under contracts DE-FG02-94ER40865 (NM) and DE-FG02-87ER40331.A008 (RL). A.T.  and P.P. acknowledge partial support from RFBR under grant No. 18-02-40086 and from the Ministry of Science and Higher Education of the Russian Federation, Project ``Fundamental properties of elementary particles and cosmology" No 0723-2020-0041. 
I.K. acknwowledges support by the Ministry of Education, Youth and Sports of the Czech Republic under grant "International Mobility of Researchers – MSCA IF IV at CTU in Prague" No.\ CZ.02.2.69/0.0/0.0/20-079/0017983.

\bibliography{ref} 
\end{document}